\documentstyle[prl,aps,epsf,twocolumn]{revtex}

\begin{document}

\draft

\preprint{DPNU-99-18, KUNS-1577}

\title{Hadronic Spectral Functions in Lattice QCD}

\author{Y. Nakahara$^{(1)}$, M. Asakawa$^{(1)}$, and  T. Hatsuda$^{(2)}$}

\address{$^{(1)}$ Department of Physics, 
Nagoya University, Nagoya 464 - 8602, Japan\\
$^{(2)}$ Physics Department, Kyoto University, Kyoto 606-8502, Japan\\
}

\date{\today}

\maketitle

\begin{abstract}
 QCD spectral functions of hadrons in the pseudo-scalar
 and vector channels are extracted from lattice Monte Carlo
 data of the imaginary time Green's functions.
 The maximum entropy method works well for this purpose,
 and the resonance and continuum structures in the spectra
 are obtained in addition to the ground
 state peaks.
\end{abstract}


 Among various dynamical quantities in quantum chromodynamics
 (QCD), the spectral functions (SPFs) of hadrons play a special role in
 physical observables  (See e.g. \cite{shuryak,negele}).
 A well-known example is the cross section of the 
 $e^+ e^-$ annihilation
 into hadrons, which can be expressed by SPF
 in the vector channel.
 SPF at finite temperature ($T$) and/or baryon density
 is also a key concept to understand the medium      
 modification of hadrons \cite{a-k}.
 The enhancement in low-mass dileptons
 observed in relativistic heavy ion collisions
 at CERN SPS \cite{ceres} is a typical
 example which may indicate a spectral shift in the medium 
 \cite{li-ko-brown}.
 
 However, the
 Monte Carlo simulations of QCD
 on the lattice, which  have been successful in measuring
 static observables \cite{lattice98},
 have difficulties in accessing the dynamical quantities 
 in the Minkowski space 
 such as SPFs and the real time correlation functions.
 This is because
 measurements on the lattice can only be carried out
 for discrete points in imaginary time.
 The analytic continuation from the imaginary time to the
 real time using the finite number of lattice data with noise 
 is highly non-trivial  and is even classified as an ill-posed
 problem. 
 
 In this paper, we make a first serious  attempt to
 extract SPFs of hadrons from lattice
 QCD data without making a priori assumptions on the spectral shape.
 For this purpose, we use the 
 maximum entropy method (MEM), which has been successfully applied for 
 similar problems in quantum Monte Carlo simulations in
 condensed matter physics,
 image reconstruction in crystallography and 
 astrophysics, and so forth  \cite{physrep,linden}.
 Due to the limitation of space, we present only the
 results for the pseudo-scalar (PS) and
 vector (V) channels at $T=0$.
 The results for other channels 
 will be given in  \cite{nextpaper}.

The Euclidean correlation function $D(\tau)$ of an 
operator ${\cal O}(\tau,\vec{x})$  and its spectral decomposition
at zero three-momentum read
\begin{eqnarray}
D(\tau ) = \int 
\langle {\cal O}^{\dagger}(\tau,\vec{x}){\cal O}(0,\vec{0})\rangle d^3 x
       =  \int_{0}^{\infty} \!\! K(\tau, \omega) A(\omega ) d\omega,
\label{KA}
\end{eqnarray}
where $\tau > 0$, $\omega$ is a real frequency, and
$A(\omega)$ is SPF
(or sometimes called the {\em image}
in this paper),
which is positive semi-definite by definition.
The kernel $K(\tau, \omega)$  is proportional to
the Fourier transform of a free boson propagator with
mass $\omega$: At $T=0$, $K(\tau, \omega)
=\exp(-\tau\omega)$. 

Monte Carlo simulation provides  $D(\tau_i)$ 
for the discrete set of points $0 \le \tau_i /a \le N_\tau$,
where $N_\tau$ is the temporal lattice size and $a$ is the 
lattice spacing.
In the actual analysis, we use data points for
$ {\tau}_{min} \le \tau_i \le {\tau}_{max} $.
From this  data set with noise, we need to reconstruct the 
continuous function $A(\omega)$ on the right hand side of
(\ref{KA}), or to make the inverse Laplace transform.
This is a typical ill-posed problem, where
the number of data is much smaller than the number of
degrees of freedom to be reconstructed.
This makes the standard 
likelihood analysis and its variants
inapplicable \cite{others} unless strong assumptions
on the spectral shape are made.
MEM is a method to circumvent this difficulty
by making a statistical inference of the most probable {\em image}
as well as its reliability \cite{physrep}.

 The theoretical basis of MEM
 is the Bayes' theorem in probability theory:
 $P[X|Y] = P[Y|X]P[X]/P[Y]$,
 where $P[X|Y]$ is the conditional probability of $X$ given $Y$.
 Let $D$ stand for Monte Carlo data with errors for a specific channel
 on the lattice and $H$ summarize all the definitions and
 prior knowledge such as  $A(\omega) \ge 0$.
 The most probable image 
 $A(\omega )$ for given lattice data is obtained by 
 maximizing the conditional probability 
 $P[A|DH]$, which, by the Bayes' theorem, is rewritten as
\begin{equation}\label{bayes_latt}
P[A|DH] \propto P[D|AH]P[A|H] ,
\end{equation}
where $P[D|AH]$ ($P[A|H]$) is called the likelihood function
 (the prior probability).

For the likelihood function, 
the central limiting theorem leads to
$P[D|AH]= Z_L^{-1} \exp (-L)$ with 
\begin{eqnarray}
L  = {1 \over 2} \sum_{i,j}
(D(\tau_i)-D^A(\tau_i))C^{-1}_{ij} (D(\tau_j)-D^A(\tau_j)),\label{chi2}
\end{eqnarray}
where $i$ and $j$ run from $ {\tau}_{min}/a $ through ${\tau}_{max}/a$.
$Z_L$ is a normalization factor given by
$Z_L = (2\pi)^{N/2} \sqrt{\det C}$ with
$N={\tau}_{max}/a - {\tau}_{min}/a+1$.
$D(\tau_i )$ is the lattice data averaged over gauge configurations 
and $D^A(\tau_i )$ is the correlation function defined 
by the right hand side of (\ref{KA}).
$C$ is an $N \times N$ covariance matrix defined by
$C_{ij}= [N_{conf}(N_{conf}-1)]^{-1}
\sum_{m=1}^{N_{conf}}
(D^{m}(\tau_i)-D(\tau_i))(D^{m}(\tau_j)-D(\tau_j))$: 
Here $N_{conf}$ is the total number of gauge configurations
and $D^{m}(\tau_i)$ is the data for the $m$-th gauge configuration.
The lattice data have generally strong correlations among
different $\tau$'s, and it is essential to take into account the
off-diagonal components of $C$.

It can be generally shown on an axiomatic basis \cite{skilling}
that, for positive distributions
such as SPF, the prior probability can be written
with parameters $\alpha$ and $m$ as
$P[A|H\alpha m]= Z_S^{-1} \exp (\alpha S)$. Here
$S$ is the Shannon-Jaynes entropy,
\begin{eqnarray}
S = \int_0^{\infty} \left [ A(\omega ) - m(\omega ) -
A(\omega)\log \left ( \frac{A(\omega)}{m(\omega )} \right ) \right ]
d\omega .
\end{eqnarray}
$Z_S$ is a normalization factor:
$Z_S \equiv  \int e^{\alpha S} {\cal D}A$.
$\alpha$ is a real and positive parameter and 
$m(\omega )$ is a real function called the default model.

In this paper, we adopt a state-of-art 
MEM \cite{physrep}, where 
the output image $A_{out}$ is
given by a weighted average over $A$ and  $\alpha$:
\begin{eqnarray}
 A_{out}(\omega)   & = &   
    \int A(\omega)  \ P[A|DH\alpha m]P[\alpha|DHm] \ {\cal D} A \ d\alpha
 \nonumber \\
    &\simeq & \int A_{\alpha}(\omega)  \ P[\alpha|DHm] \ d\alpha .
\label{final}
\end{eqnarray}
Here $A_{\alpha}(\omega)$ is obtained by
maximizing $Q \equiv \alpha S - L$ for 
a given $\alpha$, and we assume 
that $P[A|DH\alpha m] $ is
sharply peaked around $A_{\alpha}(\omega)$.
$\alpha$ dictates the relative weight of the
entropy  $S$ (which tends to fit $A$ to the default model $m$)
and the likelihood function $L$ (which tends to fit $A$ to the
lattice data). Note, however, that 
$\alpha$ appears only in the intermediate
step and is  integrated out in the final result.
We found that the weight factor $P[\alpha|DHm]$, which  
is calculable using $Q$ \cite{physrep},
is highly peaked around its maximum
$\alpha = \hat{\alpha}$ in our lattice data.
One can also study the stability of the 
$A_{out}(\omega)$ 
against a reasonable variation of $m(\omega)$.

The non-trivial part of the MEM analysis is to find the global
maximum of $Q$ in the functional space of $A(\omega)$,
which has typically 750 degrees of freedom in our case. 
We have utilized the singular value decomposition method 
to define the search direction in this functional space.
The method works successfully to find the global maximum
within reasonable iteration steps.
The technical details will be given in \cite{nextpaper}.

To check the feasibility of
the MEM procedure and to see the dependence of the 
MEM image on the quality 
of the data, we made the following test using
mock data.
(i) We start with an input  image 
$A_{in}(\omega) \equiv \omega^2 \rho_{in}(\omega)$
in the $\rho$-meson channel which simulates
the experimental $e^+e^-$ cross section.
Then we calculate $D_{in}(\tau)$ from  $A_{in}(\omega)$ using eq.(\ref{KA}). 
(ii) By taking $D_{in}(\tau_i)$ at $N$
discrete points 
and adding a Gaussian noise, we create a mock data
$D_{mock}(\tau_i)$.
The variance of the noise $\sigma (\tau_i)$
is given by $\sigma (\tau_i)= b \times D_{in}(\tau_i) \times \tau_i /a$
with a parameter $b$, which controls the noise level \cite{noise}.
(iii) We construct the output image 
$A_{out}(\omega)  \equiv \omega^2 \rho_{out}(\omega)$
using MEM with $D_{mock}(\tau_{min} \le \tau_i \le \tau_{max})$ 
and compare the result with $A_{in}(\omega)$.
In this test, we have assumed that $C$ is
diagonal for simplicity. 

In Fig.1, we show $\rho_{in}(\omega)$, and 
$\rho_{out} (\omega)$  for two
sets of parameters, (I) and  (II).
As for $m$, we choose a form 
$m(\omega) = m_0 \omega^2$ with $m_0 = 0.027$, which is
motivated by the
asymptotic behavior of $A$ in perturbative QCD,
$A(\omega \gg 1 {\rm GeV})
= (1/4 \pi^2) (1+\alpha_s / \pi) \omega^2$.
The final result is, however, insensitive to 
the variation of $m_0$ even by factor 
5 or 1/5. The calculation of $A_{out}(\omega)$
has been done by discretizing the $\omega$-space
with an equal  separation of 10 MeV between adjacent points.
This number is chosen for the reason we 
shall discuss below.
The comparison of the dashed line (set (I)) and
the dash-dotted line (set (II))
shows that increasing $\tau_{max}$ 
and reducing the noise level $b$ lead to better
SPFs closer to the input SPF. 

We have also checked that MEM can nicely reproduce
other forms of the mock SPFs.  In particular, 
MEM works very well
to reproduce not only the broad structure but also
the sharp peaks close to the delta-function as far as 
the noise level is sufficiently small.

We have then applied MEM to actual lattice data.
For this purpose, quenched lattice QCD simulations
have been done 
with the plaquette gluon action and the Wilson quark action
by the open MILC code with minor modifications \cite{milc}.
The lattice size is $20^3 \times 24$
with $\beta =6.0$, which corresponds to 
$ a = 0.0847$ fm ($a^{-1} = 2.33$ GeV),
$\kappa_c  = 0.1571$ \cite{kc}, and the
spatial size of the lattice $L_s a = 1.69 $ fm.
Gauge configurations are generated by the heat-bath and
over-relaxation algorithms with a ratio $1:4$. Each configuration
is separated by 1000 sweeps.
Hopping parameters are chosen to be
$\kappa =$ 0.153, 0.1545, and 0.1557 with $N_{conf}=161$
for each $\kappa$.
For the quark propagator, the Dirichlet (periodic)
boundary condition
is employed for the temporal (spatial) direction.
To calculate the two-point correlation functions,
we adopt a point-source at $\vec{x}=0$  and a point-sink
averaged over the spatial lattice-points
to extract  physical states with vanishing three-momentum. 
For the PS and V channels, the operators
$\bar{d} \gamma_{5} u$  and $\bar{d} \gamma_{\mu} u$ ($\mu = 1,2,3$) 
are chosen, respectively.
We use data at $1 \le \tau_i/a  \le 12$ 
to remove the noise at the Dirichlet boundary.
To avoid the known pathological behavior of the eigenvalues of 
$C$ \cite{physrep}, we take $N_{conf} \gg N$.

We define SPFs for the PS and V channels as
\begin{equation}
 A(\omega) = \omega^2 \rho_{_{PS},_{V}}(\omega) ,
\end{equation}
so that $\rho_{_{PS,V}}(\omega \rightarrow {\rm large}) $ approaches
a finite constant as predicted by  perturbative QCD.
For the MEM analysis,
we need to discretize the $\omega$-integration in (\ref{KA}).
Since $\Delta \omega$ (the mesh size) $\ll 1/\tau_{max}$
should be satisfied  to suppress the discretization error,
we take $\Delta \omega$ = 10 MeV.
$\omega_{max}$ (the upper limit for the $\omega$ integration)  
should be comparable to the maximum
available momentum on the lattice:
$\omega_{max} \sim \pi /a \sim 7.3$ GeV.
We have checked that larger
values of $\omega_{max} $ do not change 
the result of $A(\omega)$ substantially, while smaller
values of $\omega_{max} $ distort the high energy end of the
spectrum. The dimension
of the image to be reconstructed  is $N_{\omega} \equiv
\omega_{max}/\Delta \omega \sim 750$,
which is in fact much larger than the maximum number of 
Monte Carlo data  $N = 25$.
In the MEM analysis presented in this paper,
the continuum kernel $K = \exp(-\tau \omega)$ is used.
We have also carried out analysis based on the lattice kernel
$K^{lat}$. A comparison of the two cases will be given
in \cite{nextpaper} in detail.

In Fig.2 (a) and (b), we show the reconstructed images for
each $\kappa$. In these figures, we have used $m = m_0 \omega^2$ with
$m_0 = 2.0 (0.86)$ for PS (V) channel motivated by
the perturbative estimate of $m_0$ 
(see eq.(\ref{cont-V}) and the text below).
We have checked that 
the result is not sensitive, within the statistical
significance of the image, to the variation
of $m_0$ by factor 5 or 1/5. 
The obtained images have a common structure:
the low-energy peaks corresponding to
$\pi$ and $\rho$, and the broad structure in the high-energy
region.  From the position of the pion peaks in Fig.2(a), we extract 
$\kappa_c = 0.1570(3)$, which is
consistent with $ 0.1571 $ \cite{kc} determined from the
asymptotic behavior of $D(\tau)$. 
The mass of the $\rho$-meson in the chiral limit
extracted from the peaks in Fig.2(b) reads
$m_{\rho}a = 0.348(15)$. This  is also consistent with 
$m_{\rho}a = 0.331(22) $  \cite{kc}
determined by the asymptotic behavior.
Although our maximum value of the
fitting range $\tau_{max}/a =12$ 
marginally covers the asymptotic limit in $\tau$, we 
can extract reasonable masses for $\pi$ and $\rho$.
The width of $\pi$ and $\rho$  in Fig.2
is an artifact due to the statistical errors of the
lattice data. In fact, in the quenched approximation, there is no room 
for the $\rho$-meson to decay into two pions.

As for the second peaks in the PS and V channels,
the error analysis discussed in Fig.3 shows that
their spectral ``shape" does not have much  statistical
significance, although the existence of the
non-vanishing spectral strength is significant.
Under this reservation, we fit the position of the 
second peaks and made linear
extrapolation to the chiral limit with 
the results, $m^{2nd}/m_{\rho} = 1.88(8)
(2.44(11))$  for the PS (V) channel.
These numbers should be compared with the 
experimental values:
$m_{\pi(1300)}/m_{\rho} = 1.68$,
and $m_{\rho(1450)}/m_{\rho} = 1.90$
or $m_{\rho(1700)}/m_{\rho} = 2.20$.

One should remark here that, 
in the standard two-mass fit of $D(\tau)$, the mass
of the second resonance is highly sensitive to the 
lower limit of the fitting range, e.g., 
$m^{2nd}/m_{\rho} = 2.21(27) (1.58(26))$  for $\tau_{min}/a = 8 (9)$
in the $V$ channel with $\beta=6.0$ \cite{kc}.
This is because the contamination from the short distance
contributions from $\tau < \tau_{min}$ is not
under control in such an approach.
On the other hand, MEM does not suffer from this difficulty 
and can utilize the full
information down to $\tau_{min}/a=1$.
Therefore, MEM opens
a possibility of systematic study of higher resonances 
with lattice QCD data.

As for the third bumps in Fig.2, the spectral ``shape" 
is statistically not significant as is discussed in Fig.3,
and they should rather be considered a part of the
perturbative continuum instead of a single resonance.
Fig.2 also shows that SPF decreases substantially
above 6 GeV; MEM automatically detects
the existence of the momentum cutoff on the lattice $\sim \pi/a$.
It is expected that MEM with the data on finer lattices leads to larger
ultraviolet cut-offs in the spectra.
The  height of the asymptotic form of the
spectrum at high energy is estimated as 
\begin{equation}
 \rho_{_V}(\omega \simeq 6 {\rm GeV})
 = {1 \over 4 \pi^2} \left ( 1 + {\alpha_s \over \pi} \right )
 \left ( {1 \over 2\kappa  Z_{_V}} \right )^2 \simeq 0.86 .
\label{cont-V}
\end{equation}
The first two factors are the $q \bar{q}$
continuum expected from  perturbative QCD.
The third factor contains  the non-perturbative
renormalization constant for the lattice composite operator.
We adopt 
$Z_{_V} = 0.57$ determined from the two-point functions at
$\beta$ = 6.0 \cite{mm86} together with $\alpha_s = 0.21$
and $\kappa = 0.1557$.
Our estimate in eq.(\ref{cont-V}) 
is  consistent with the high energy part of the spectrum
in Fig.2(b) after averaging over $\omega$.
We made a similar estimate for the PS channel using
$Z_{_{PS}} = 0.49 $ \cite{shi} and obtained
$\rho_{_{PS}}(\omega \simeq 6 {\rm GeV}) \simeq 2.0$. This is
also consistent with Fig. 2(a).
We note here that 
an independent analysis of the imaginary time
correlation functions \cite{negele}
also shows that 
the lattice data at short distance is dominated by 
the perturbative continuum.

Within the MEM analysis, one can study the statistical 
significance of the reconstructed image by the following
procedure \cite{physrep}.
Assuming that $P[A|DH\alpha m]$ has a Gaussian distribution
around the most probable image $\hat{A}$, we estimate the error 
by the covariance of the image, 
$- \langle (\delta_A \delta_A Q)^{-1} \rangle_{A=\hat{A}}$,
where $\delta_A$ is a functional derivative and
$\langle \cdot \rangle $ is an average over
a given energy interval.
The final error for $A_{out}$ is obtained 
by averaging the covariance over $\alpha$ with
a weight factor $P[\alpha |DHm]$.
Shown in Fig.3 is  the MEM image in the V channel for $\kappa= 0.1557$
with errors obtained in the above procedure. 
The height of each horizontal bar is
$\langle\rho_{out}(\omega)\rangle$
in each $\omega$ interval.
The vertical bar indicates the error of 
$\langle\rho_{out}(\omega)\rangle$.
The small error for the lowest peak in Fig.3
supports our identification of the peak with $\rho$. 
Although the existence of the non-vanishing
spectral strength of the 2nd peak and 3rd bump
is statistically significant, their spectral ``shape''
is either marginal or insignificant.
Lattice data with better quality are called for
to obtain better SPFs.

In summary, we have made a first serious attempt to reconstruct
SPFs of hadrons from lattice QCD data.
We have used MEM, which allows us to study SPFs without
making a priori assumption on the spectral shape.
The method works well for the mock data. 
Even for the lattice data,
the method produces resonance  and continuum-like structures
in addition to the ground state peaks.
The statistical significance of the image has been also analyzed. 
Better data with finer and larger lattice will
produce better images with smaller errors, and our
study should be considered a first attempt towards this goal.
We have not made the chiral extrapolation of SPFs in this paper,
since we have found that neither the
direct extrapolation of the MEM image $A_{out}(\omega)$ 
nor the extrapolation of $D(\tau)$ and $C$
works in a straightforward manner.
We leave this as an open problem. 
From the physics point of view, the spectral change at finite
temperature is by far the important problem. This is 
currently under investigation.

We appreciate MILC collaboration for their open codes
for lattice QCD simulations, which have enabled this research.
Our simulation was carried out on a Hitachi 
SR2201 parallel computer at Japan Atomic Energy Research Institute.
M. A. (T. H.) was partly supported by Grant-in-Aid for Scientific
Research No. 10740112 (No. 10874042) of the Japanese Ministry of Education,
Science, and Culture. T. H. was also supported by Sumitomo
Foundation (Grant No. 970248).


\begin{figure}
\epsfxsize=8.2cm
\centerline{\epsfbox{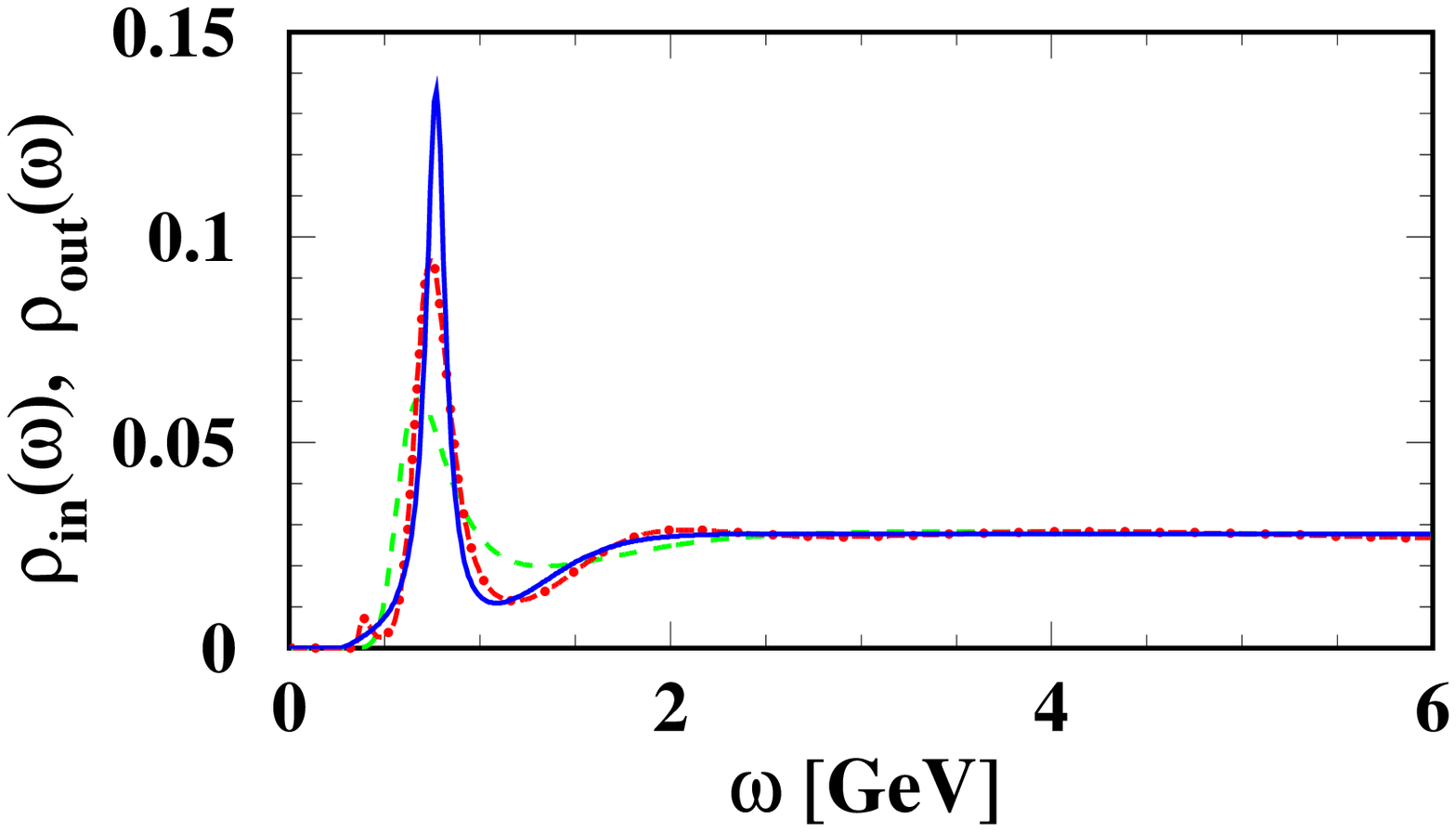}}
\vskip0mm
\caption{
 The solid line is $\rho_{in}(\omega)$.
 The dashed line and dash-dotted line are $\rho_{out}(\omega)$
 obtained with  parameter set (I) $a = 0.0847$ fm, 
 $1 \le \tau /a \le 12$, $b=0.001$ 
 and set (II) $a = 0.0847$ fm, 
 $1 \le \tau /a \le 36$, $b=0.0001$, respectively. }
\label{fig1}

\vskip4mm
\epsfxsize=8.2cm
\centerline{\epsfbox{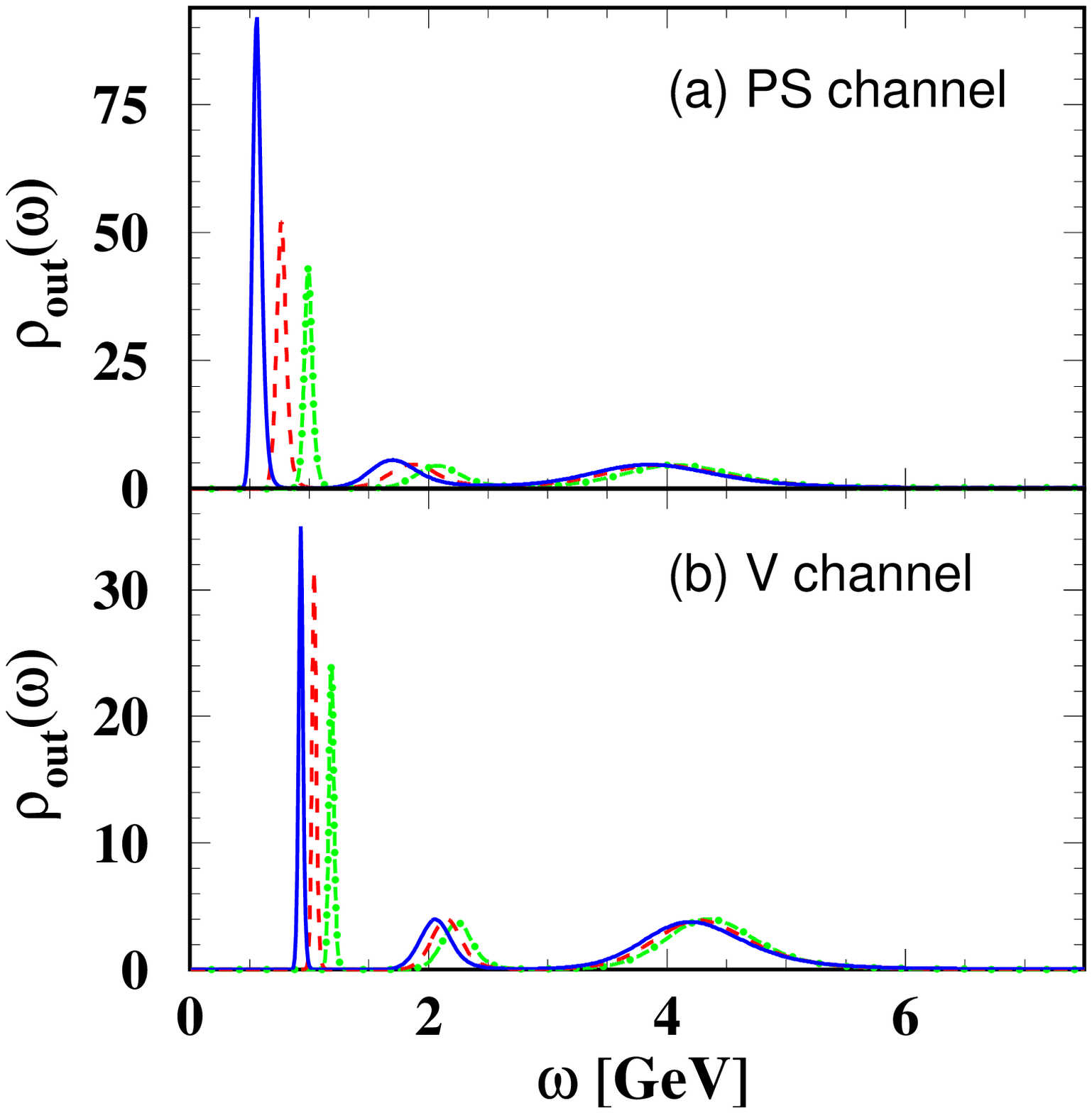}}
\vskip2mm
\caption{
	Reconstructed image $\rho_{out}(\omega)$ for
the PS (a) and V (b) channels. The solid, dashed, and dash-dotted
lines are for $\kappa=$ 0.1557, 0.1545, and 0.153, respectively.
For the PS (V) channel, $m_0$ is taken
to be 2.0 (0.86). $\omega_{max}$ is 7.5 GeV 
in this figure and Fig.3.}
\label{fig2}

\vskip4mm
\epsfxsize=8.2cm
\centerline{\epsfbox{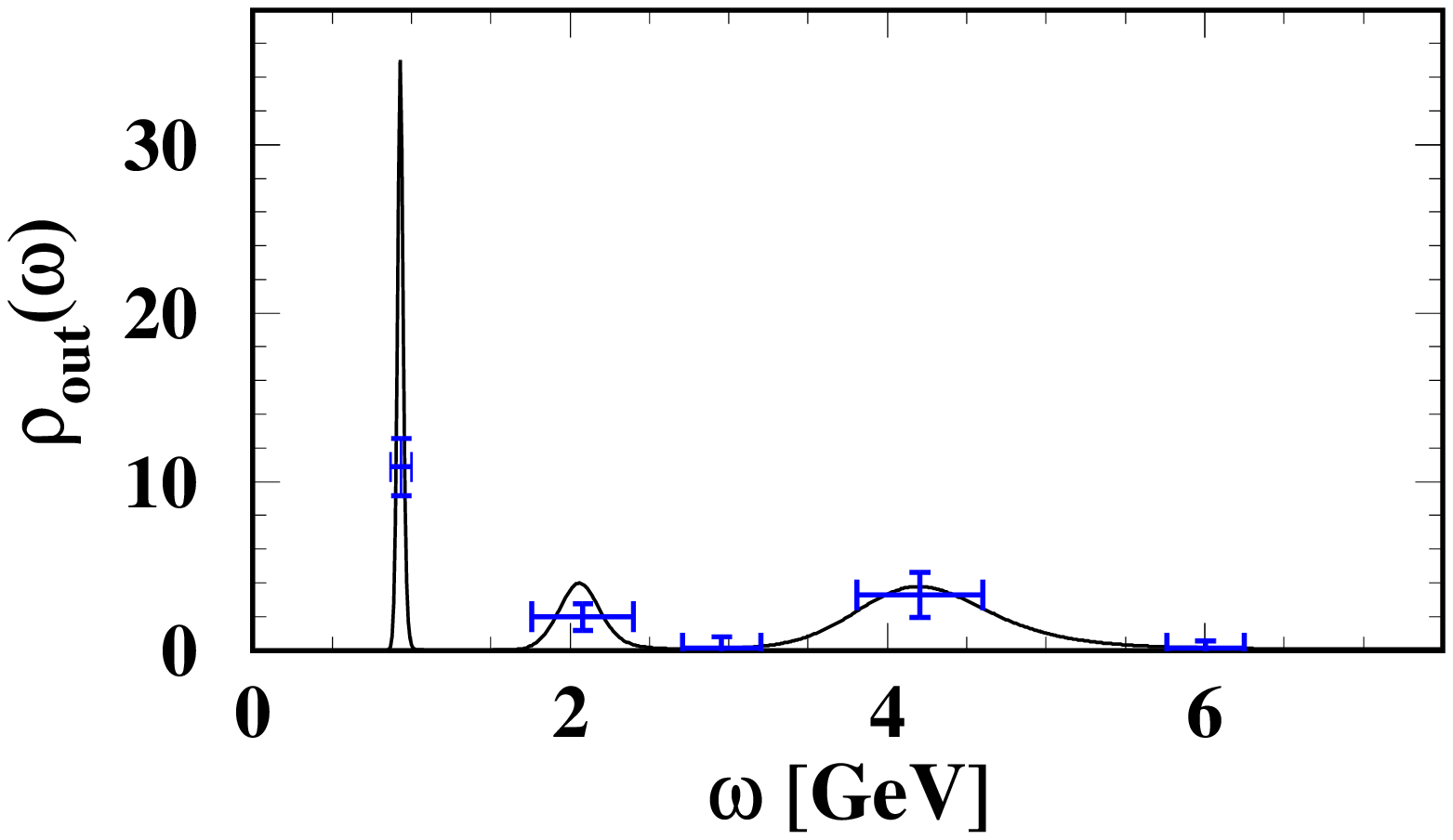}}
\vskip2mm
\caption{
	$\rho_{out}(\omega)$ in the $V$ channel 
 for $\kappa=$ 0.1557 with error attached.}
\label{fig3}

\end{figure}


\begin{thebibliography}{99}
\bibitem{noise}
        This formula is motivated by our lattice QCD data.
\end{thebibliography}
\end{document}